\documentclass[twocolumn]{revtex4}

\usepackage{amsmath, amssymb, graphics}
\usepackage{hyperref}
\usepackage{graphicx}
\usepackage{array}

\newcommand{\eps}{\epsilon}

\newcommand{\ket}[1]{\lvert #1 \rangle}

\newcommand{\bra}[1]{\langle #1 \rvert}

\newcommand{\abs}[1]{\lvert#1\rvert}

\newcommand{\PR}{\emph{Phys.~Rev.}~}
\newcommand{\PRL}{\emph{Phys.~Rev.~Lett.}~}
\newcommand{\RMP}{\emph{Rev.~Mod.~Phys}~}
\newcommand{\NJP}{\emph{New Journal of Physics}~}

\newcommand{\hc}{\text{h.c.}}

\renewcommand{\i}{\mathrm{i}}
\renewcommand{\eps}{\epsilon}

\renewcommand{\exp}[1]{\mathrm{e}^{#1}}

\renewcommand{\d}{\mathrm{d}}

\begin{document}
\title{On Rabi oscillations between Bloch bands}
\author{Patrick Pl\"otz}
\affiliation{Institut f\"ur Theoretische Physik, Universit\"at Heidelberg, 69120 Heidelberg, Germany}
\begin{abstract}
We study Rabi oscillations between the bands of an arbitrary biased superlattice in a tight-binding model. We reduce the problem to an equation of Whittaker--Hill type and, in absence of any known solutions in closed form, discuss different approximations to describe the oscillations between the Bloch bands. We identify regimes of weak and strong inter-band coupling and compare predictions for these Rabi oscillations to numerical results.
\end{abstract}

\maketitle
\section{Introduction}
The motion of electrons in a periodic potential under a constant external force is one of the fundamental phenomena of solid-state physics and treated in many text books~\cite{Kittel}. Bloch oscillations~\cite{BlochOsci} and transitions between energy bands via Landau--Zener tunneling~\cite{Zener} are important effects discussed in this context. They have been studied in many different contexts, like electrons in semiconductor superlattices~\cite{SCSL}, optical waves in photonic lattices~\cite{Photons}, and Bose--Einstein condensates in optical lattices~\cite{BECs}. The latter allow a particularly high degree of control over various system parameters.

In the present work, we consider a two-band tight-binding model with an additional tilting force and study the force-induced coupling between the two Bloch bands. This has been an active area of research for quite some time, although mainly focused on electronic motion in superlattices~\cite{Abumov, Zhao, Rotvig, Fukuyama, BlochZenerTheo}. For ultra-cold atoms in optical lattices~\cite{BlochZwergerReview,morsch}, nearly closed two-band systems are also realisable using different techniques as described, e.g., in~\cite{WeitzMinibands}. Besides the possibility of experimental realisation, a closed two-band model is also interesting as a simple model system. Here, we discuss the two lowest energy bands of an optical lattice $\tilde V(x) = V_0\cos (2 k_L x)$, with the wave vector of the optical lattice $k_L = 2\pi/\lambda_L$. All energies are measured in units of the recoil energy $E_R\equiv \hbar^2k_L^2/(2m)$ and we set $\hbar =1$. This has the advantage, that all parameters of the model Hamiltonian depend only a single quantity, the depth of the optical lattice $V_0$, and can easily be computed~\cite{TMW1}.

\begin{figure}[h]
\center
	\includegraphics[width = 0.9\linewidth,clip]{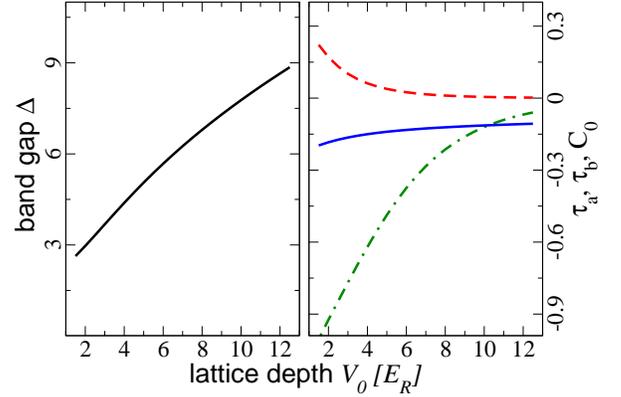}
	\caption{Hamiltonian parameters as a function of the lattice depth $V_0$. \emph{Left panel:} Average band gap $\Delta$. \emph{Right panel:} Hopping coefficients $\tau_a$ (dashed), $\tau_b$ (dashed-dotted), and interband coupling constant $C_0$ (solid).}\label{fig:parameters}
\end{figure}

The two-band Hamiltonian in the tight-binding approximation reads
\begin{equation}\label{eq:H0}
	\begin{split}
		\mathcal H  =  - \frac{\tau_a}{2} \sum_{l} \left( \ket{a_{l+1}}\!\bra{a_l} + \hc \right) - \frac{\tau_b}{2} \sum_{l} \left( \ket{b_{l+1}}\!\bra{b_l} + \hc \right) \\
		 +  \sum_{l} \eps_l^- \ket{a_l}\!\bra{a_l} + \eps_l^+ \ket{b_l}\!\bra{b_l} + V\left( \ket{a_l}\!\bra{b_l}  + \hc \right)
	\end{split}
\end{equation}
with $\eps_n^{\pm} = \pm\Delta/2 + nF$. Here, $\tau_{a(b)}$ denote the hopping strength in band $a(b)$, $\eps_l^{\pm}$ the on-site energies, and $V=C_0 F$ the local coupling between the energy bands. In a realisation with optical lattices, the parameters $\Delta, C_0, \tau_a, \tau_b$ depend on the depth of the lattice $V_0$,  fig.~\ref{fig:parameters} shows this dependence of the Hamiltonian parameters for different values of $V_0$. In the following, we consider the external Stark force $F$ as a free parameter and our focus will lie on the oscillations between the two Bloch bands.

\section{Momentum Space Analysis}
\begin{figure}[t]
\center
	\includegraphics[width = 0.9\linewidth,clip]{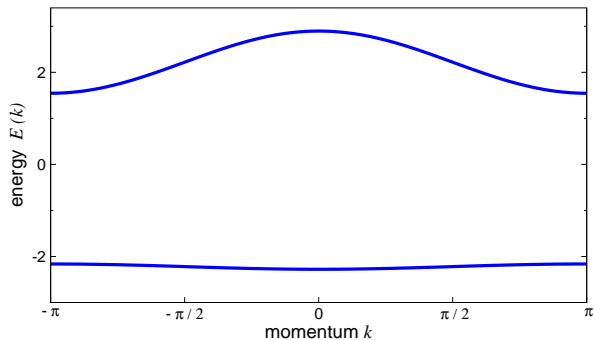}
	\caption{The two energy bands (solid line) of Eq.~(\ref{eq:Hkspace}) in momentum space at $t=0$. A non-zero force $F$ leads to periodic motion within the energy bands. Parameters are for $V_0 = 4$: $\Delta = 4.39, \tau_a = 0.062, \tau_b = -0.62$.} \label{fig:bands}
\end{figure}

We start by changing to Fourier space in two steps: first we remove the force term in the time-dependent Schr\"odinger equation by performing a gauge transformation and in a second step we will switch to momentum space. The applied force breaks the translational invariance of the system which can be restored by changing to the interaction-picture with respect to the external force~\cite{FB}, such that the Stark term is absent but momentum $\hat p$ is replaced by $\hat p-Ft$ and the system is explicitly time-dependent. We introduce Fourier components
	\begin{equation}
		a(k) = \sum_l \exp{i lk}a_l	\quad \& \quad b(k) = \sum_l \exp{i lk}b_l
	\end{equation}
and obtain from the original non-interacting two-band model the following time-periodic two-level Hamiltonian~\cite{Zhao}
	\begin{equation}\label{eq:Hkspace}
	 H (k,t) =	\binom{ -\frac{\Delta}{2} - \tau_a \cos(k+Ft) \qquad\quad V \quad\qquad}{ 
			\qquad\quad V \qquad\quad \frac{\Delta}{2} - \tau_b \cos(k+Ft) }
	\end{equation}
The spectrum at fixed time is shown in fig.~\ref{fig:bands}.
Transforming $\tilde a(k,t)  = a(k,t) \text{exp}[i(-\frac{\Delta}{2}t - \tau_a \int_0^t \cos(k + Ft')\d t')]$ and $\tilde b(k,t)  = b(k,t) \text{exp}[i(+\frac{\Delta}{2}t - \tau_b \int_0^t \cos(k + Ft')\d t')]$, one arrives at a purely off-diagonal Schr\"odinger equation
	\begin{equation}\label{eq:Schroedinger}
		i\partial_t \binom{\tilde a(k,t)}{\tilde b(k,t)} = \left( \begin{array}{cc} 0 & V\exp{-i\phi(k,t)} \\ 
			V\exp{i\phi(k,t)} & 0 \end{array}\right)\binom{\tilde a(k,t)}{\tilde b(k,t)}
	\end{equation}
where $\phi(k,t) = \Delta \cdot t - \Delta x[\sin(k+Ft) - \sin(k)]$ is the phase between the two Bloch bands and $\Delta x = (\tau_a-\tau_b)/F$. Where the pure existence of two energy bands allows phenomena like Rabi oscillations, it is the non-trivial phase difference caused by $\Delta x \neq 0$, that gives rise to interesting, non-trivial phenomena for the time-evolution in both bands to be discussed below. Please note, that our model is slightly different from a tight-binding superlattice, as e.g. in~\cite{BlochZenerTheo}, since we include a local on-site coupling of the bands. Let us discuss the special case $\tau_a = \tau_b$ first. The non-linear time dependence vanishes and we can write the system of differential equations as a linear differential equation of second order. We differentiate the second equation and insert it into the first one, arriving at the simple equation
$\ddot{ \tilde a }+ i\Delta \dot{\tilde a} +V^2 \tilde a = 0.$ This is easily solved and one finds Rabi-like oscillations 
\begin{equation}\label{eq:Rabi}
	|a(k,t)|^2 = \frac{V^2}{{\Delta^2+4V^2}}\sin^2 (\tilde\Delta \cdot t/2)
\end{equation}
with frequency $ \tilde\Delta=\sqrt{\Delta^2+4V^2}$ and an amplitude much smaller than unity for the parameters chosen here. The same procedure can be applied for the general problem, but the differential equation to be solved has now time-dependent  coefficients
	\begin{equation}\label{eq:second_order}
		0 = \ddot{ \tilde a }+ i\Big[\Delta +(\tau_a-\tau_b)\cos(k+Ft)\Big] \dot{\tilde a} +V^2 \tilde a.
	\end{equation}
We make an ansatz $a = y\, \text{exp}[\i\Omega(t)]$ and choose 
\begin{equation}
	\dot\Omega = -\frac{1}{2}\left[ \Delta +(\tau_a-\tau_b)\cos(k+Ft)\right], 
\end{equation}
to remove the second term in eq.~(\ref{eq:second_order}). We finally arrive at 
\begin{equation}
	\begin{split}
	 0= 2\ddot{ y}+ y\Big[2 V^2-\i F(\tau_a-\tau_b)\sin(k+F t) \\
		 -3\big(\Delta+(\tau_a-\tau_b)\cos(k+Ft)\big)^2\Big]. 
	\end{split}
\end{equation}
This equation is of the Whittaker--Hill type, described for instance in~\cite{hillequation}, and solutions in closed form are not known for the general case. For the rest of this work, we will study different approximate solutions of the problem, i.e., of eq.~(\ref{eq:H0}) and~(\ref{eq:Schroedinger}). 

\begin{figure}[t]
\center
	\includegraphics [angle = 0, width=0.9\linewidth]{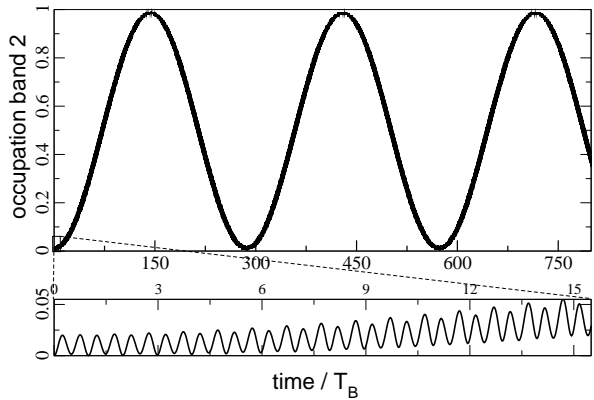}
	\caption{Occupation of band 2 as function time in resonance. We observe long period and large amplitude Rabi oscillations with smaller and much faster oscillations on top. Parameters as in fig.~\ref{fig:bands} and $F=2.2207$.}
	\label{fig:Rabires}
\end{figure}

\begin{figure}[t!]
\center
	\includegraphics [angle = -90, width=0.9\linewidth]{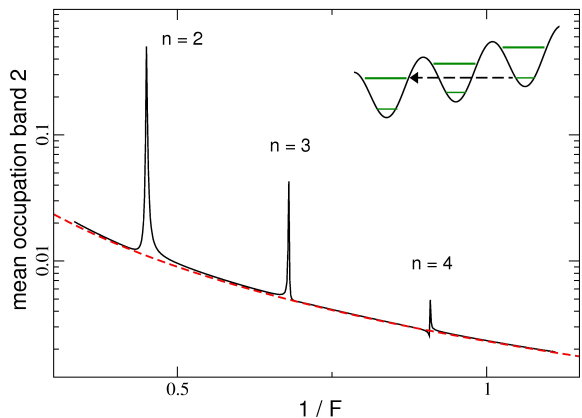}
	\caption{Long time averages of the occupation of band 2 as function of the inverse force. There is a clear signature of resonant behaviour when the band gap equals an integer multiple of the force. The dashed line shows the average occupation of band 2 as expected from eq.~(\ref{eq:Rabi}) and the inset schematically shows an $n=2$ resonance. Parameters as in fig.~\ref{fig:bands}.}
	\label{fig:resonances}
\end{figure}

A formal solution of eq.~(\ref{eq:Schroedinger}) is given by
	\begin{equation}
		\binom{\tilde a(k,t)}{\tilde b(k,t)}  =  \mathcal T\text{exp}\left[-\i\int_0^t \tilde H(k,t') \d t'\right] \binom{\tilde a(k,0)}{\tilde b(k,0)}, 
	\end{equation}
where $\mathcal T$ denotes time-ordering. Following Zhao \emph{et al.}~\cite{Zhao}, it can be solved approximately by dropping the time-ordering (valid to first order in $V$).  Choosing occupation in the lower band only as initial state, i.e. $a(k,0) = 1$ and $b(k,0) = 0$, one obtains for the occupation of the upper band
	\begin{equation}
		|b(k,t)|^2 = \sin^2\left| V \int_0^t \exp{i\phi(k,t')} \d t'\right|, 
	\end{equation}
with $\phi(k,t) = \Delta \cdot t - \Delta x[\sin(k+Ft) - \sin(k)]$ as above. Unfortunately, this integral can not be expressed in simple terms of elementary functions, but has to be treated in different levels of approximation or numerically. To obtain an idea of the physics contained in the above expression, we stick to some simple approximation and will find signs of resonances in this system. Writing $\exp{i \phi} = \cos \phi +i \sin \phi$ and using the addition theorems, we find integrands of the form $\cos(\Delta x \sin t)$ and $\sin(\Delta x \sin t)$ which are not accessible. We therefore expand the functions in Fourier series 
	\begin{align*}
		\cos (\Delta x \sin t)  = J_0(\Delta x) + \sum_{n=1}^{\infty}2J_{2n}(\Delta x)\cos(2nFt) \\
		\sin (\Delta x \sin t)  = \sum_{n=0}^{\infty}2J_{2n+1}(\Delta x)\sin\left((2n+1)Ft\right)
	\end{align*}
where $J_l(x)$ denotes the ordinary Bessel function of order~$l$. Taking only the first terms, the integrals can be solved analytically and expressions of the following form ($n = 1,2$) appear
	\begin{equation*}
		|b(k,t)|^2 \propto \sin^2\left|\sin^n(\Delta x /n)\; \frac{\sin [(\Delta - nF)\,t]}{\Delta  nF}\right|.
	\end{equation*}
The higher orders in the expansion lead to similar terms. Thus, a truncated Fourier expansion indicates the existence of resonances in the occupation of the upper band for $\Delta\approx nF$, that is for values of the external force where the tilt between sites of different bands equals the band gap. To test this expectation, we evolved an initial state with $a(k,0) = 1$ and $b(k,0)=0$ in time for a resonant value and the resulting occupation of the upper band $|b(t)|^2$ is shown in fig.~\ref{fig:Rabires}. We observe small Rabi oscillations with a high frequency, which are described by eq.~(\ref{eq:Rabi}), and on top of these are Rabi oscillations of large amplitude and long period $T\gg T_B$. The latter oscillations only take place near resonances $\Delta\approx nF$.
\\
We did similar time evolutions for various values of the external force $F$ and studied the long time average of $|b(t)|^2$. The result is depicted in fig.~\ref{fig:resonances}. It clearly demonstrates the existence of resonances in our system. 
The occupation of the upper band is usually very small, of the order of a few percent, but shows sharp resonances for $nF \approx \Delta$. A closer numerical study (not shown here) shows a Lorentzian shape for the lower order resonances and the possibility of more complicated, Fano-like shapes for the narrower higher order resonances. The non-resonant oscillations are well described by eq.~(\ref{eq:Rabi}), compare also fig.~\ref{fig:resonances}. Below we will derive an effective model for the system in resonance that allows a quantitative prediction of the resonance width. Similar resonant behaviour has been observed in extensive numerical studies of the occupation of minibands for electronic motion in superlattices~\cite{Abumov}. 

\section{System in Resonance}
\begin{figure}[h]
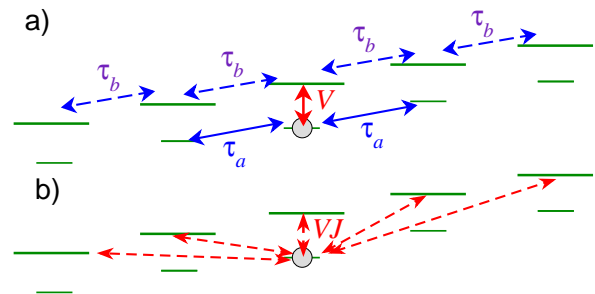

\center
	\includegraphics[width=0.9\linewidth]{noninter1.eps}
	\includegraphics[width=0.9\linewidth]{noninter2.eps}
	\caption{Graphical representation of the two-band Hamiltonian: a) in the original basis, eq.~(\ref{eq:H0}), and b) in the transformed basis, eq.~(\ref{eq:Htransformed}).\label{fig:twoband_noninter}} 
\end{figure}

Having found resonances in the system, we are now going study this resonant behaviour in more detail. Starting from the original Hamiltonian in coordinate space, eq.~(\ref{eq:H0}), we apply a  transformation known~\cite{Fukuyama} to diagonalise the single band problem
\begin{subequations}\label{eq:trafo}
	\begin{eqnarray}
		\ket{\alpha_n}  = \sum_l J_{l-n}(\tau_a/F) \ket{a_l} \\
		\ket{\beta_n}  = \sum_l J_{l-n}(\tau_b/F) \ket{b_l} 
	\end{eqnarray}
\end{subequations}
where we defined $x_{a(b)} := \tau_{a(b)}/F$ and $\Delta x = x_a - x_b$. Using properties of Bessel functions, we arrive at 
	\begin{equation}\label{eq:Htransformed}
	\begin{split}
		\mathcal H = \sum_l \Big[ \eps_l^{-} \ket{\alpha_l}\!\bra{\alpha_l} +\eps_l^{+}\ket{\beta_l}\!\bra{\beta_l} \;+ \qquad\qquad\\ V \sum_n J_{l-n}(\Delta x)\left( \ket{\alpha_l}\!\bra{\beta_{l+n}}  + \hc \right) \Big].
	\end{split}	
	\end{equation}
We obtain a coupling between any two sites of the two different bands, weighted by Bessel functions. This means the coupling between different and possibly remote sites, which is originally mediated by on-site coupling and subsequent hoppings, can be considered a higher order process in the original basis. But in the transformed basis eq.~(\ref{eq:Htransformed}), this coupling is now direct with a strength modified by the factor $J_{l-n}(\Delta x)$. This is schematically depicted in fig.~\ref{fig:twoband_noninter}.
Interestingly, the same matrix representation can be found by application of Floquet theory~\cite{ShirleyPaper} to the time-periodic Hamiltonian eq.~(\ref{eq:Schroedinger}). One has to solve the larger but time-independent eigenvalue problem for the Floquet operator $\mathcal H_F$ having matrix elements
	\begin{equation*}
		(\mathcal H_F)_{\alpha\gamma}^{n,k} = H_{\alpha \gamma}^{n-k} + nF \delta_{\alpha\gamma}\delta_{nk}, \quad n,k \in \mathbb{Z}; \; \alpha,\gamma = a,b.
	\end{equation*}
This is based on expanding the Hamiltonian in a Fourier series with Fourier components $H^l$. In our case these Fourier components are 2$\times$2 matrices and read explicitly
\begin{subequations}
	\begin{eqnarray}
		H^{0} & = & \left( \begin{array}{cc} -\Delta/2  & V \\ V & \Delta/2  \end{array}\right)\\
		H^{2n} & = & V J_{|2n|}(\Delta x)\left( \begin{array}{cc} 0& 1 \\ 1 & 0 \end{array}\right) \\
		H^{2n+1} & = & \text{sgn}(n)V J_{|2n+1|}(\Delta x)\left( \begin{array}{cc} 0& -1 \\ 1 & 0\end{array}\right), 
	\end{eqnarray}
\end{subequations}
with the signum function $\text{sgn}(x)$. The obtained Floquet Hamiltonian $\mathcal H_F$ is the same as obtained earlier in eq.~(\ref{eq:Htransformed}). For the special case $\Delta x=0$, i.e. $\tau_a = \tau_b$, the model eq.~(\ref{eq:Htransformed}) can be solved exactly as before, cf. eq.~(\ref{eq:Rabi}). The argument of the Bessel functions is zero for $\Delta x =0$ and only on-site coupling of the two bands remains. 

When the system is exactly in resonance, there is a degeneracy between two energy levels of the two bands and the most relevant coupling in the transformed basis, is the direct one between these two levels. Taking only this coupling, we arrive at an effective Rabi problem~\cite{Nakamura}. In a resonance of order $m$, this reduces to a simple 2$\times$2-matrix: 
	\begin{equation}
		\mathcal H_{\text{eff}}^{(m)} = \left(\begin{array}{cc} \epsilon^{-}_{N+m} & VJ_m(\Delta x) \\ V J_m(\Delta x) & \epsilon_N^+ \end{array}\right)
	\end{equation}
The occupation of the upper band follows a simple Rabi formula
	\begin{equation}\label{eq:Rabitransformed}
		\abs{b(t)}^2 = \frac{4V^2 J_m^2}{(mF-\Delta)^2 + 4V^2J_m^2}\; \sin^2\left(\Omega_{\text{res}} t\right), 
	\end{equation}
with a Rabi frequency for the resonant oscillations given by $\Omega_{\text{res}}= \frac{1}{2}\sqrt{( \epsilon^{-}_{N+m}- \epsilon^{+}_{N}) +4V^2J_m^2}\approx \abs{V J_m(\Delta x)}$. This frequency is much smaller than the Bloch frequency $\omega_B=2\pi/T_B$ and the corresponding periods are given by
\begin{equation}\label{eq:Tres}
	T_{\text{res}}^{(m)} = \frac{F}{2\abs{V J_m(\Delta x)}}\;T_B.
\end{equation}
Thus, the system in resonance exhibits oscillations between the two bands of an amplitude close to unity and on a time scale much larger than the direct on-site oscillations.
\begin{figure}[ht]
\center
	\includegraphics[width = 0.9\linewidth,clip]{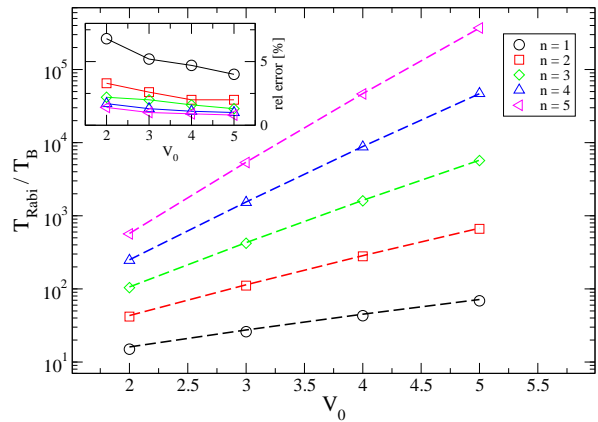}
	\caption{Periods of resonant oscillation exactly in resonance as predicted by mapping on a two-level Rabi problem (dashed lines) and compared to periods from numerical time evolution in perturbation theory (data points). The inset shows the relative error between the two values.} \label{fig:Trabi}
\end{figure} 
In fig.~\ref{fig:Trabi} we compare this predictions to numerical simulations for different values of the parameters in the Hamiltonian, eq.~(\ref{eq:H0}), and different orders of resonance. We observe very good agreement. Additionally, the inset shows the relative error between the predicted and numerically observed values. It is approximately 5~\% for the widest resonance $n=1$, but smaller than 3~\% for higher resonances and slightly decreasing with increasing potential depth. 

We want to compare these results to nearly degenerate perturbation theory in the original basis. Here, the resonances of different order have to be treated separately. Starting with the first order resonance ($\Delta = F$), one finds in second order degenerate perturbation theory a Rabi frequency of $\Omega = \frac{1}{2}\abs{V (\tau_b-\tau_a)/\Delta}$ reproducing the leading order result of $\Omega_{\text{res}}^{(m=1)} = \abs{VJ_1(\Delta x)}$. For the second order resonance ($\Delta = 2F$), one obtains in third order degenerate perturbation theory $\Omega = \frac{1}{2}\abs{V (\tau_b^2+\tau_a^2)/(8F^2)}$, slightly differing from the leading order behaviour of the result in the transformed basis. Higher order resonances can be computed in a similar fashion, but the resonant basis~(\ref{eq:trafo}) is clearly a better starting point for further calculations. Indeed, considering the mean occupation of the upper band close to a resonance, eq.~(\ref{eq:Rabitransformed}), we find a Breit-Wigner shape for the resonances as also seen numerically~\cite{Abumov}, namely
	\begin{equation}
		\langle \abs{b(t)}^2\rangle (1/F) =  \frac{1}{2}\frac{\Gamma_m^2}{(m/\Delta-1/F)^2+\Gamma_m^2}, 
	\end{equation}
where $\Gamma_m = 2\abs{VJ_m(\Delta x)}/(F\Delta)$ and the width of the resonances is accordingly given by $2\Gamma_m$. 
Comparing to numerical simulations (not shown here), the effective model presented here gives a good estimate for the magnitude of the resonance width. In particular, the width decreases very quickly with the order of the resonance or the depth of the optical lattice, since
\begin{equation}
	2\Gamma_m \approx \frac{4\abs{C_0} }{\Delta} \frac{m^m}{2^mm!} \left(\frac{\tau_a-\tau_b}{\Delta}\right)^m,
\end{equation}
where the factor $m^m/2^mm!$ is of order one (for $m = 1, \ldots, 10$), but since $\left(\tau_a-\tau_b\right)\ll\Delta $, the last factor decreases very rapidly. We do not compare these values with results from numerical calculations in more detail here. Higher accuracy can be obtained by degenerate perturbation theory in the resonant basis. 

\section{Summary}

We studied oscillations in a two-band tight-binding model. We could reduce the problem of solving the time-dependent Schr\"odinger equation to a Whittaker--Hill equation and concentrated on different physical regimes for the inter-band oscillations. We found and discussed resonances in the system and derived an effective model for the system in resonance allowing a quantitative characterisation of the resonances themselves and the time-scales of inter-band oscillations.

\medskip

\emph{This work was supported within the framework of the Excellence Initiative by the German Research Foundation (DFG) through the Heidelberg Graduate School of Fundamental Physics (grant number GSC 129/1) and the Klaus Tschira Foundation.}

\end{document}